\documentclass[a4paper,10pt]{article}
\usepackage[utf8]{inputenc}
\usepackage{jheppub}
\usepackage{amsmath}
\usepackage{slashed}
\usepackage{axodraw}
\usepackage[T1]{fontenc}
\usepackage{comment}
\usepackage{hyperref}
\usepackage{braket}
\usepackage{graphicx}

\title{ \boldmath Scalar Dark Matter in Scale Invariant Standard Model}

\author[a]{Karim Ghorbani}
\author[b]{Hossein Ghorbani}

\affiliation[a]{Physics Department, Faculty of Sciences, Arak University, Arak 38156-8-8349, Iran}
\affiliation[b]{Institute for Research in Fundamental Sciences (IPM), 
 School of Particles and Accelerators, P.O. Box 19395-5531, Tehran, Iran}

\emailAdd{kghorbani@ipm.ir}
\emailAdd{pghorbani@ipm.ir}

\date{}
\abstract{We investigate single and two-component scalar dark matter scenarios in classically scale invariant 
standard model which is free of the hierarchy problem in the Higgs sector. We show that despite the very restricted 
space of parameters imposed by the scale invariance symmetry, both single and two-component scalar dark matter models
overcome the direct and indirect constraints provided by the Planck/WMAP observational data and the LUX/Xenon100 experiment.
We comment also on the radiative mass corrections of the classically massless scalon that plays a crucial role in our study.}

\begin{document}
\maketitle

\section{Introduction}

The Standard Model (SM) is the most successful theory of interacting fundamental particles known 
so far. Nevertheless, there
are theoretical and observational shortcomings which are left unanswered up to now. Some important examples of such 
drawbacks are the hierarchy in the Higgs sector \cite{Weinberg:1975gm,Gildener:1976ai} and the problem of 
dark matter (DM) and dark energy (see \cite{Adam:2015rua,Hinshaw:2012aka} for the latest Planck/WMAP results). 
Therefore it seems 
that a theory beyond the standard model is inevitable. Among other extensions to the standard model, 
the minimal supersymmetry 
standard model (MSSM) has drawn a lot of attentions as it is capable in addressing 
the above mentioned problems (see e.g. \cite{Gondolo:2004sc,Jungman:1995df} for the DM issue in SUSY  and 
see for instance \cite{Brax:2011bh} and the references therein 
for the DE study in SUGRA). Despite the broad consensus on the fact that SUSY might be observed 
in the experiments
but the latest results of the Large Hadron Collider (LHC) experiments, 
CMS and ATLAS \cite{Aad:2015baa,Khachatryan:2015exa} 
in run-I show no evidence for any s-particle. Although it is still soon to consider the 
MSSM theory excluded before the forthcoming results of the LHC run-II come up, however there is
already enough motivation to think about the alternative theories. 

In the SM scheme the natural is that the electroweak scale and the Planck scale be of the same order. 
We observe instead that the scale of the weak interactions is much smaller than the GUT or the Planck scale.
What causes this problem is the introduction of a mass scale in the electroweak sector, that is the Higgs mass. 
From the quantum corrections the physical Higgs mass should be much bigger than what we observe 
in the experiments unless a delicate fine-tuning takes place in the theory. 
An example of such fine-tuning is the cancellation of the quantum corrections of the Higgs mass in the MSSM. 

A different way to avoid the hierarchy problem is setting to zero the 
tree-level quadratic Higgs mass term in the standard model lagrangian \cite{Meissner:2006zh,Foot:2007iy}. 
The resulting theory is usually called the scale invariant standard model (SISM) or
the conformal standard model (CSM). But if the Higgs particle is {\it massless} in the SM how the spontaneous
symmetry breaking can occur without the Higgs mass term? 
The answer is that in the quantum level 
the Higgs scalar gains a small mass from the radiative corrections that is a conformal anomaly that
breaks the scale invariance. This conformal anomaly is the means for spontaneously breaking the
electroweak symmetry. 
It was demonstrated first by Coleman and E. Weinberg (CW) \cite{Coleman:1973jx} 
that an abelian gauge theory possessing a massless scalar can undergo the spontaneous symmetry breakdown in 
the vacuum expectation value of the massless scalar through the 
radiative corrections. 

This idea was implemented for the standard model
by Gildener and S. Weinberg (GW) \cite{Gildener:1976ih} 
who argued that 
the SISM with $n$ scalars consists a number of ``heavy Higgs'' and one light scalar named ``scalon''. 
In order for the standard model to include the heavy Higgs boson with mass around $125$ GeV observed by the LHC
in July 2012 the model must possess at least three scalars. For some recent work on dark matter in the framework 
of the scale invariant extension of the standard model see \cite{Guo:2014bha,Karam:2015jta}.

The goal of the current paper is to examine if the scale invariant standard model with a number of
weakly coupled massless scalars can account for the WIMP candidate produced via the freeze-out mechanism. 
The two scalar extension to SISM, where both scalar bosons 
take non-zero {\it vevs} giving the correct values for Higgs mass, turn out to be not an appropriate DM freeze-out
model as the light scalon is unstable in the lack of the $\mathbb{Z}_2$ symmetry. Nevertheless
further scalar fields in  
the scale invariant lagrangian can play the role of the DM if we keep only the $\mathbb{Z}_2$ symmetric terms   
involving the DM candidates and require that the DM scalar takes zero {\it vev}. 

In this paper we examine the SISM with one heavy Higgs, one light scalon as the mediator 
between SM and DM sectors, and additional one and two 
scalars as single DM and two-component DM candidates respectively. 
Both single scalar DM and two-component scalar DM are consistent with the direct and indirect
constrains. 

The paper is arranged as the following. In the next section we elaborate the scale 
invariant standard model and introduce the scalar dark matter extension to that. In particular
we will see that the radiative correction to scalon mass given in eq. (\ref{m_s}) is crucial in finding
a consistent model of dark matter. In section \ref{prob} we test the DM scenarios with the 
WMAP and Planck observational data, and LUX and Xenon100 direct experiments. In two subsections 
\ref{SDM} and \ref{DDM}
we study the validity of the single and two-component scalar models against such bounds. 
In the last section we summarize the results and discuss about the possible scale 
invariant fermionic extension and explain why it disparages the fermionic DM candidate 
in SISM.

\section{Extending Scale Invariant Standard Model}\label{model}

Standard model is classically scale invariant provided that the Higgs mass term is absent. However  
it is possible for such a theory to gain mass through an anomaly. 
It was shown for the first time in the seminal work of Coleman and E. Weinberg (CW) \cite{Coleman:1973jx} 
that the massless scalar electromagnetic theory is spontaneously broken through radiative corrections where
both the gauge vector and the 
scalar field in the theory become massive. The scale invariant standard model with massless Higgs 
at the classical level can become massive by the same mechanism 
as worked out by Gildener and S .Weinberg (GW) \cite{Gildener:1976ih}. 
In the version of the SM studied by GW there is no Higgs mass term. The only term remaining in the Higgs potential 
is the quartic term $\lambda_H \left( H^\dagger H\right)^2$. However, in the absence of the mass term 
$m_H^2  H^\dagger H$, the theory can not admit a non-zero vacuum expectation value. To cure this problem  
 they add instead a number of scalar fields in the quartic form $\lambda_{ijkl}\Phi_i\Phi_j\Phi_k\Phi_l$ to the 
 {\it classically massless SM}. They show that through the radiative corrections in the GW theory 
 there exist a number of ``heavy'' Higgs 
with a mass comparable to intermediate gauge bosons together with a ``light'' scalar which they dub scalon. 
We take this light scalar (scalon) as a mediator coupled both to the heavy Higgs in the SISM and to the dark sector. 
In the dark sector the additional scalars are such that the new terms preserve the scale invariance so they are in the
quartic form as introduced by GW. Furthermore
the DM scalars enjoy the $\mathbb{Z}_2$ to be stable. The lagrangian under the above circumstances takes the following form,
\begin{equation}\label{lag1}
 \mathcal{L}_{\text{SISM}}=\mathcal{L}'_{\text{SM}}-V(H,s,\varphi_i)\,,
\end{equation}
where $\mathcal{L}'_{\text{SM}}$ is the massless Higgs SM (standard model lagrangian 
without the usual Higgs potential term). 
The potential term $V(H,s,\varphi_i)$ is defined as
\begin{equation}\label{pot}
 V(H,s,\varphi)=\frac{\lambda_H}{4}(H^\dagger H)^2+\frac{\lambda}{2}s^2(H^\dagger H)
 +\frac{\lambda_s}{4}s^4+ \frac{1}{2} s^2  \sum_i{\lambda_i \varphi_i^2} 
 + \frac{1}{4}\sum_i{\lambda_{\varphi_i}} \varphi_i^4 \,,
\end{equation}
where $H$, $s$, and $\varphi_i$ are respectively the doublet Higgs, the scalon and DM scalars. In the current work we 
consider only $i=1,2$ cases. At the minimum of the
potential (\ref{pot}) the Higgs field $H$ and the scalar $s$ take non-vanishing vacuum expectation 
values $v_H$ and $v_s$ and the 
DM scalars $v_{\varphi_i}$ takes vanishing {\it vev}, $v_{\varphi_i}=0$.

To apply the same approach as GW \cite{Gildener:1976ih} we need to find a flat direction in some RG scale $\Lambda$ 
in the scalar fields configuration along which the potential 
(\ref{pot}) vanishes. From now on we use only real singlet scalar $h$, the only 
component of the complex Higgs doublet which is left after the symmetry breaking.  
We can describe the configuration of the real scalar fields $h$ and $s$ in terms of the 
spherical coordinates of angles $\mathbf{n}\equiv \left( {\hat{n}_h=\cos{\theta},
\hat{n}_s=\sin{\theta}} \right )$  
and a radial field $\phi$. Then $h=\hat{n}_h \phi$ and $s=\hat{n}_s \phi$. Let 
$\mathbf{\bar{n}}\equiv \left( \cos{\alpha},\sin{\alpha} \right )$ 
be along the flat direction for some RG scale 
$\mu=\Lambda$, then we have 
\begin{equation}\label{flatdir}
 V(\mathbf{\bar{n}}\phi)=0 \Rightarrow \cos^2{\alpha}= 
 \frac{\lambda_H -\lambda \pm \sqrt{\lambda^2-\lambda_H \lambda_s}}
 {\lambda_H-2\lambda+\lambda_s}\,,
\end{equation}
therefore there are two flat directions for the potential (\ref{pot}).
We could pick only one flat direction by choosing, 
\begin{equation}\label{angle}
 \lambda^2-\lambda_H \lambda_s=0  \hspace{.5cm} \Rightarrow \hspace{.5cm} \cos^2{\alpha}=  \frac{\lambda_H}
 {\lambda_H-\lambda}\,.
\end{equation}
The local minimum of the tree-level potential along the 
flat direction occurs at the {\it vev}s $v_H$ and $v_s$
by setting the first derivative of the potential (\ref{pot}) 
to zero at the special $\alpha$ given in (\ref{angle}), 
\begin{equation}\label{min}
\frac{\partial V}{\partial x}  \Big\rvert_{\mathbf{\bar{n}}\braket{\phi}} =0 \Rightarrow \hspace{1cm}
\lambda_H v_H^2 =-\lambda v_s^2 \hspace{1cm} \lambda_s v_s^2=- \lambda v_H^2\,,
\end{equation}
where $x=h,s, \varphi_i$. Using eq. (\ref{min}) the tree-level mass 
matrix is easily driven at the {\it vevs},
\begin{equation}\label{offmass}
 \mathbf{M^2_{\text{tree}}}\equiv \frac{\partial^2 V}{\partial x \partial y}  \Big\rvert_{\mathbf{\bar{n}}\braket{\phi}}=
\begin{pmatrix}
\mathbf{M}_{\text{tree}}^2(h,s) & 0\\
0 & \lambda_i v_s^2  \\ 
\end{pmatrix}\,,
\end{equation}
where $x,y=h, s, \varphi_i$. The mass matrix (\ref{offmass}) is a diagonal $3\times 3$ matrix if $i=1$ and a diagonal
$4\times 4$ matrix if $i=2$ with, 
\begin{equation}\label{massHs}
\mathbf{M}_{\text{tree}}^2(h,s)=2\lambda_H v_H^2
\begin{pmatrix}
  1 & - v_H / v_s \\
  -v_H / v_s  & v_H^2 / v_s^2\\
\end{pmatrix}.
\end{equation}
After diagonalizing the mass matrix (\ref{offmass}) the tree-level mass eigenvalues 
for all scalars in the theory are obtained as the following, 
\begin{equation}\label{mass}
 m_H^2= 2(\lambda_H - \lambda) v_H^2, \hspace{1 cm} m_s^2=0, 
 \hspace{1cm} m_{\varphi_i}^2= -\frac{\lambda_H \lambda_i}{\lambda} v_H^2\,.
\end{equation}
The masses $m_H$ and $m_s$ in eq. (\ref{mass}) become diagonal entries of the mass matrix
(\ref{massHs}) if we make a rotation in 
$(h,s)$ space, 
\begin{equation}
\begin{pmatrix}
 h \\
 s \\ 
\end{pmatrix} \rightarrow
\begin{pmatrix}
\cos\alpha &  - \sin\alpha \\
 \sin\alpha  &  \cos\alpha \\ 
\end{pmatrix} 
\begin{pmatrix}
h  \\
 s  \\ 
\end{pmatrix} \,,
\end{equation}
with $\alpha$ being the angle given in eq. (\ref{angle}).
As long as the scalon $s$ is massless it can be shown that the elastic scattering cross section
of DM off nuclei becomes drastically large and the model is immediately excluded by the direct detection experiments. 
The singlet scalar $s$ however receives a small mass 
along the flat direction via the radiation corrections.  
The effective potential then reads \cite{Coleman:1973jx},
\begin{equation}\label{Veff}
 \delta V(\bar{\mathbf{n}}\phi)\equiv V^{\text{1-loop}}_{\text{eff}}(\bar{\mathbf{n}}\phi)
 =A(\bar{\mathbf{n}}) \phi^4 + 
 B(\bar{\mathbf{n}}) \phi^4 \log{\frac{\phi^2}{\Lambda^2}}\,,
\end{equation}
where $A(\bar{\mathbf{n}})$ and $B(\bar{\mathbf{n}})$ are dimensionless coefficients, 
\begin{equation}\label{An}
\begin{split}
 A(\bar{\mathbf{n}})=\frac{1}{64\pi^2 v_{\phi}^2} 
 \bigg[ m_h^4 \left( -\frac{2}{3} 
 + \log{\frac{m_h^2}{v_\phi^2}}\right)
 + m_{\varphi_i}^4 \left( -\frac{2}{3} 
 + \log{\frac{m_{\varphi_i}^2}{v_\phi^2}}\right)
 + 6 m_W^4 \left( -\frac{5}{6} 
 + \log{\frac{m_W^2}{v_\phi^2}}\right) \\
 + 3 m_Z^4 \left( -\frac{5}{6} 
 + \log{\frac{m_Z^2}{v_\phi^2}}\right)
 -12 m_t^4 \left( -1
 + \log{\frac{m_t^2}{v_\phi^2}}\right) 
 \bigg]\,,
 \end{split}
\end{equation}
and 
\begin{equation}\label{Bn}
 B(\bar{\mathbf{n}})=\frac{1}{64\pi^2 v_{\phi}^4} 
 \left( m_h^4 +m_{\varphi_i}^4 + 6m_W^4 + 3m_Z^4 -12 m_t^4 \right)\,.
\end{equation}
In eqs. (\ref{An}) and (\ref{Bn}) $v_\phi$ stands for the {\it vev} of the radial field $\phi$,
and the factors behind the quartic masses are the number of degrees of freedom for the fields
appearing in the 
loop. It has been shown in \cite{Gildener:1976ih} that the mass correction to the classically 
massless scalar $s$ is given by 
\begin{equation}
\delta m_s^2 =
\frac{d^2\delta V(\bar{\mathbf{n}}\phi)}{d\phi^2}\Big\rvert_{\braket{\phi}}
=12 v_\phi^2 \left( A +\frac{7}{6} B  + B  \log{\frac{v_\phi^2}{\Lambda^2}} \right)\,,
\end{equation}
 where the minimization condition of the effective potential (\ref{Veff}) at the {\it vev} $v_\phi$,
 \begin{equation}
  \Lambda = v_\phi  \exp{\left( \frac{A}{2B}+\frac{1}{4} \right)}\,,
 \end{equation}
leads to 
\begin{equation}\label{m_s}
 \delta m_s^2 = 2 B v_\phi^2=-\frac{\lambda}{32\pi^2 m_H^2} 
 \left( m_H^4+ m_{\varphi_i}^4+ 6m_W^4 + 3m_Z^4 -12 m_t^4 \right)\,,
\end{equation}
where eq. (\ref{min}) and $v_\phi^2=v_H^2+v_s^2$ have been used. The tree-level potential
in the flat direction can be expressed in terms of 
the couplings $\lambda_H, \lambda$, $\lambda_i$, $\lambda_{\varphi_i}$ and the Higgs {\it vev} $v_H$,

\begin{equation}
\begin{split}
 V(h,s,\varphi_i)= \frac{1}{2}m_H^2 h^2 +\frac{1}{2} m_{\varphi_i}^2 \varphi_i^2
 + \left( \lambda_H + \lambda \right) \sqrt{1-\frac{\lambda}{\lambda_H}} v_H h^3
 + \frac{1}{4} \frac{\left( \lambda_H + \lambda \right)^2}{\lambda_H} h^4 \\
 + \left( \lambda_H  + \lambda \right) \sqrt{-\frac{\lambda}{\lambda_H}} h^3 s 
 + 2\sqrt{-\lambda \left (\lambda_H -\lambda \right) } v_H h^2 s
 - \lambda h^2 s^2 \\
 + \frac{\lambda_H \lambda_i v_H}{\sqrt{-\lambda (\lambda_H-\lambda)}} s \varphi_i^2
 - \frac{\lambda_i \lambda_H v_H}{\sqrt{\lambda_H (\lambda_H-\lambda)}} h \varphi_i^2
 + \frac{\sqrt{-\lambda \lambda_H}\lambda_i}{\lambda_H-\lambda} s h \varphi_i^2 \\
 +\frac{1}{2} \frac{\lambda_H \lambda_i}{\lambda_H-\lambda} s^2 \varphi_i^2
 -\frac{1}{2} \frac{\lambda_i \lambda}{\lambda_H-\lambda} h^2 \varphi_i^2
 +\frac{1}{4} \lambda_{\varphi_i} \varphi_i^4\,.
 \end{split}
\end{equation}
where for the single DM $i=1$ and for the two-component DM $i=1,2$.
\section{Direct and Indirect Probes}\label{prob}
In this section we check the validity of the single and two-component scalar dark matter models introduced in 
section \ref{model} 
against  the direct experiments and the indirect observational data. We have utilized the package 
micrOMEGAs 
 \cite{Belanger:2013oya,Belanger:2014vza} 
for numerically computing the relic density and the DM-nucleon 
elastic scattering cross section.
\subsection{Single Dark Matter}\label{SDM}
The case of single scalar dark matter is the simplest dark matter scalar model 
in the scale invariant standard model. 
There are three
types of scalars involved here. The SM Higgs scalar $h$, the mediator scalar $s$ 
which gives mass to the Higgs $h$ if taking non-zero expectation value $v_s$.
The latter cannot be a DM candidate in the freeze-out scenario as it decays into other particles in SM. 
Now further scalars that we add to the theory provided that they take zero {\it vev} and interact 
only with the scalar $s$ will be stable and hence play the role of dark matter particles. 

The parameters used in the theory (\ref{lag1}) are the set 
$\{\lambda_H, \lambda,\lambda_s, \lambda_1,\lambda_{\varphi_1},\alpha,v_s^2 \}$.
Evidently there is no mass parameter for the Higgs field due to the scale invariance.
Taking into account the eqs. (\ref{angle}), (\ref{min}) and (\ref{mass}) the free independent parameters reduce to 
$\{\lambda, \lambda_1\}$ and $\lambda_{\varphi_1}$ where 
the parameter $\lambda_{\varphi_1}$ do not enter into calculations at tree level in perturbation theory.

\begin{figure}
\begin{minipage}{0.36\textwidth}
\includegraphics[width=\textwidth,angle =-90]{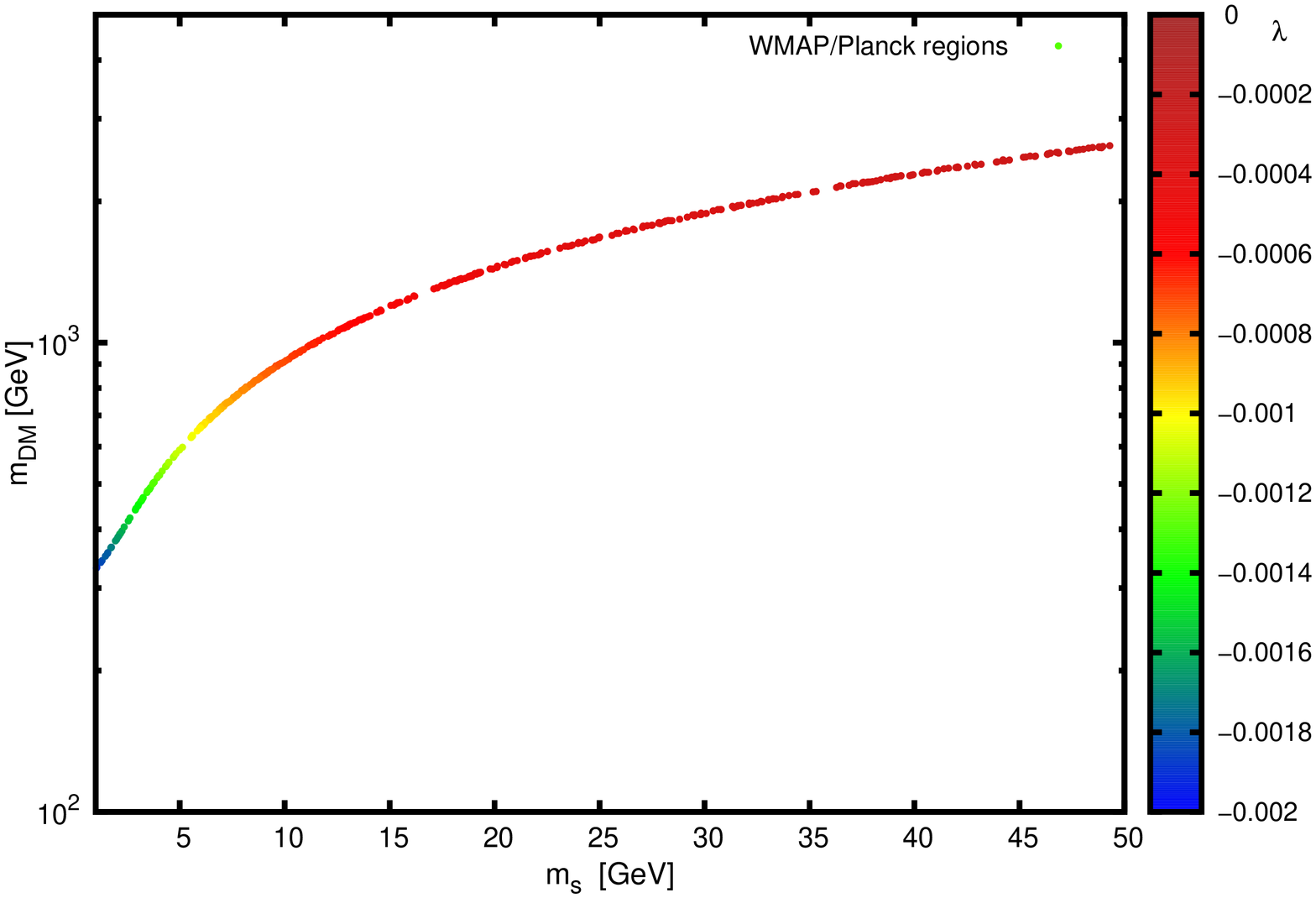}
\end{minipage}
\hspace{2.4cm}
\begin{minipage}{0.36\textwidth}
\includegraphics[width=\textwidth,angle =-90]{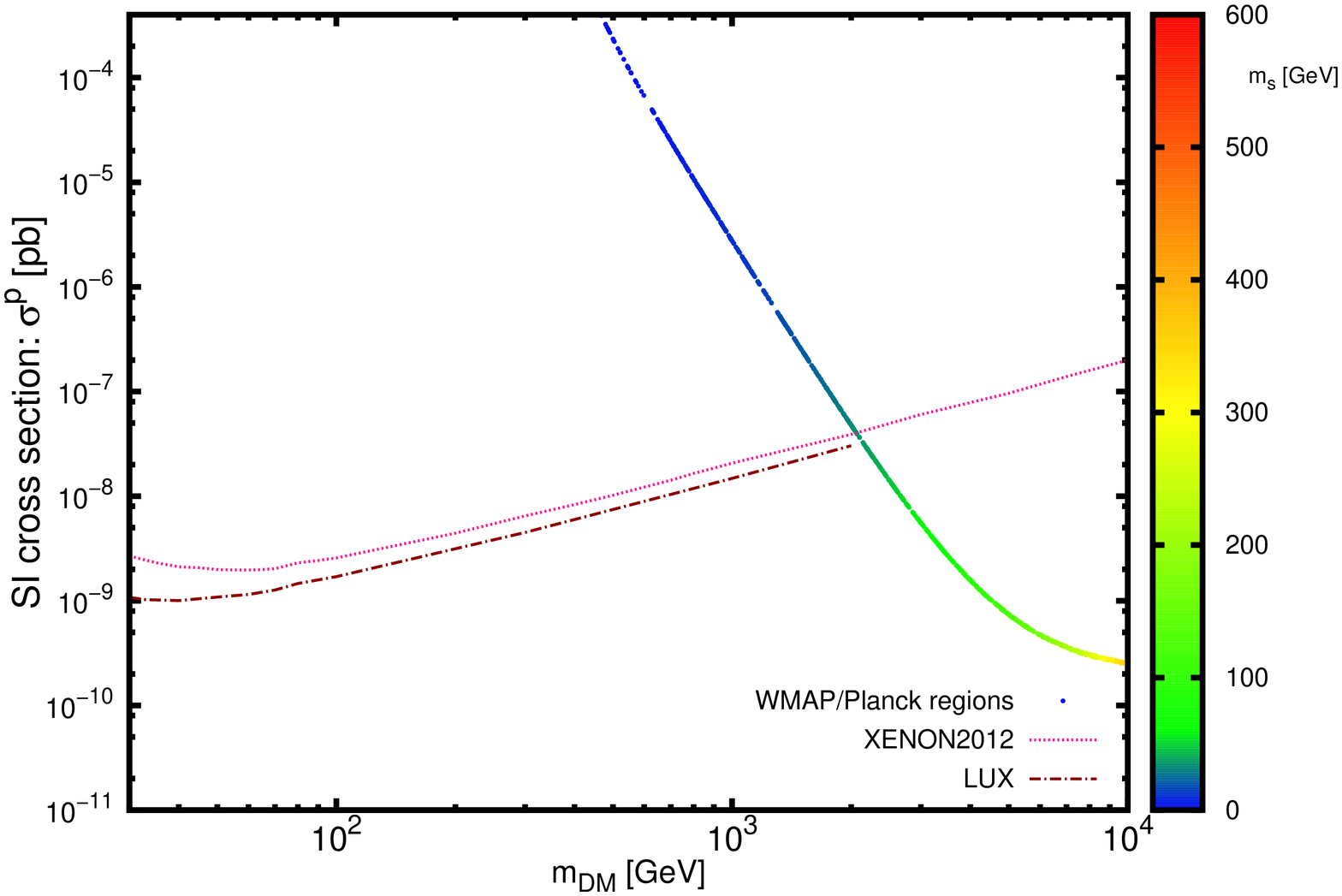}
\end{minipage}
\caption{Direct and Indirect probes for single scalar DM in the SISM: {\it Left}) 
The allowed DM mass constrained from DM relic abundance measured by Planck/WMAP against the scalon mass, {\it Right})
The allowed DM mass constrained by Planck/WMAP for relic density and by Xenon100/LUX for the elastic
scattering cross section of DM off the proton.}
\label{singleDM}
\end{figure}

To evaluate the relic density for the single scalar DM scenario we need to solve the Boltzmann
differential equation for the  time evolution of number density $ n_{\varphi_1}$,
\begin{equation}\label{boltz}
\frac{dn_{\varphi_1}}{dt}+3Hn_{\varphi_1}= -\braket{\sigma_{\text{ann}} 
v_{\text{rel}}} \left[ n_{\varphi_1}^2-\left( n^{\text{EQ}}_{\varphi_1}\right )^2 \right] \,,
\end{equation}
where $H$ is the Hubble expansion rate, the $\braket{}$ means thermal averaging, $\sigma_{\text{ann}}$ 
denotes the dark matter annihilation cross section and $v_{\text{rel}}$ is the relative velocity 
(for more details on Boltzmann equation see e.g. \cite{Dodelson:2003ft,Ghorbani:2015baa}). 
The stability condition puts already some constraints on the space 
of parameters. From eqs. (\ref{min}) and (\ref{mass}) we find that 
$v_s^2=-m_H^2/2\lambda-v_H^2$. Now fixing the Higgs mass to $m_H = 125$ GeV and the Higgs {\it vev},
 $v_H=246$ GeV and using the fact that $v_s^2>0$, it turns out that $-0.128<\lambda<0$.
 Then from $m_H^2=2(\lambda_H-\lambda)v_H^2$ in (\ref{mass}) we get $\lambda_{H} = \lambda + 0.128$.
 Finally in eq. (\ref{mass}) we have $m_{\varphi_i}^2=-\lambda_H \lambda_i v_H^2 /\lambda$. Substituting 
 $m_{\varphi_i}^4$
into eq. (\ref{m_s}) and putting the masses of $m_H, m_Z, m_W$ and $m_t$ we obtain from $\delta m_s^2>0$ that
$\lambda_i > -1.65 \lambda /\lambda_H$.
 We see that the scale symmetry not only decreases the number of independent parameters but
 also constrains quite strongly the parameter space. 
We have solved the Boltzmann equation (\ref{boltz}) by scanning over the allowed values 
of the couplings $\lambda$ and $\lambda_1$ and kept only the couplings that give 
the correct relic abundance $ 0.1172 < \Omega_{\text{DM}} h^2 < 0.1226 $ for dark matter measured 
accurately by WMAP and Planck. 
Both the mediator mass given in (\ref{m_s}) and the mass of the DM i.e. $m_{\text{DM}}\equiv m_{\varphi_1}$  in (\ref{mass}) 
are related directly to the couplings $\lambda$ and $\lambda_1$. In Fig. \ref{singleDM} the dependence of the $m_{\text{DM}}$
on the mediator mass $m_s$ for the constrained parameter space from the relic density is plotted. It is seen from 
the left plot in Fig. \ref{singleDM} that the DM mass grows for smaller coupling $\lambda$ which in turn leads
to greater mediator mass $m_s$.

We can check our model as well with the direct detection measurements studying the elastic scattering cross section of 
the dark matter off the nuclei used in the experiments. The scalar DM in our model interacts with the quarks
through only the Higgs portal which is a mixing of the scalars $s$ and $h$ here. The Feynman diagram describing the
DM-quark interaction $\varphi_1 q \rightarrow  \varphi_1 q$ is a tree-level diagram drawn in Fig. 5 in  
\cite{Ghorbani:2014gka}. The effective potential for such an interaction is given by

\begin{equation}\label{Leff}
\mathcal{L}_{\text{eff}} =\alpha_q \varphi_1 \varphi_1 \bar q q \,,
\end{equation}
where the effective coupling $\alpha_q$ is 
\begin{equation}
 \alpha_q= m_q \frac{2 \lambda_H \lambda_1}{\lambda_H - \lambda} (\frac{1}{m_s^2}+\frac{1}{m_H^2}) \,.
\end{equation}
It is a good approximation if we consider only the zero momentum transfer in the DM-nucleon scattering. In this limit the
quark currents are replaced by the  nucleonic currents and the spin-independent (SI) elastic 
scattering cross section reads
\begin{equation}
 \sigma_{\text{SI}}^{\text{N}}=\frac{\alpha_N^2 \mu_N^2}{\pi m_{\text{DM}}^2}\,,
\end{equation}
where $\alpha_N$ is a factor depending on the effective coupling in eq. (\ref{Leff}), the mass of the quarks, the 
quarks scalar form factors, and the mass of the nucleon which e.g. in the Xenon experiments is the xenon mass
(see for instance \cite{Ellis:2008hf,Belanger:2008sj,Crivellin:2013ipa} for more details). 
We have calculated such elastic scattering for the parameter space which is already restricted by the 
relic density bounds imposed by Planck and WMAP. Despite the very narrow parameter space we deal with
in the current model we observe that still for dark matter masses heavier than around $2$ TeV 
we have a viable parameter space that respect both the Planck and the Xenon100 bounds. 
In fact for $m_{\text{DM}}\gtrsim 2$ TeV not only the 
the bounds by LUX and Xenon100 are respected but even the forthcoming bounds by Xenon1T might not 
exclude the model. This result is obvious from the right plot in Fig. \ref{singleDM}. 
We should emphasize the role of the non-zero although small scalon mass, in obtaining
acceptable results for an only 2-dimensional shrieked parameter space. It is clearly seen from the right panel
in Fig. \ref{singleDM} that when the scalon mass goes to zero the scattering cross section grows very fast. 
A small change to scalon mass from e.g. $0$ to $1$ GeV reduces the cross section for about $15$ orders of 
magnitudes!

\subsection{Two-component Dark Matter}\label{DDM}

Now in addition to scalars $h$ and $s$ we consider two scalars $\varphi_1$ and $\varphi_2$ as DM particles with again
vanishing {\it vev}. This will be a {\it two-component} example of dark matter models. 
The set of parameters are enlarged compared to that of single scalar dark matter and is 
$\{\lambda_H, \lambda,\lambda_s, \lambda_1,\lambda_2,\lambda_{\varphi_1},\lambda_{\varphi_2},
\alpha,v_s^2 \}$. The independent parameters that inter in the calculations are the 
set $\{\lambda, \lambda_1,\lambda_2\}$. Notice that both DM particles $\varphi_1$ and $\varphi_2$ are stable;
non of them decays into the other or to SM particles. The time evolution of each DM scalar is evaluated by two
independent Boltzmann equations, 
\begin{equation}\label{2boltz1}
 \frac{dn_{\varphi_1}}{dt} = -3Hn_{\varphi_1} -\braket{ \sigma^{11}_{\text{ann}}v^{11}_{\text{rel}}}
 \left [n_{\varphi_1}^2-\left( n^{\text{eq}}_{\varphi_1}\right )^2 \right]\,,
  \end{equation}
\begin{equation}\label{2boltz2}
 \frac{dn_{\varphi_2}}{dt} = -3Hn_{\varphi_2} -\braket{ \sigma^{22}_{\text{ann}}v^{22}_{\text{rel}}}
 \left [n_{\varphi_2}^2-\left( n^{\text{eq}}_{\varphi_2}\right )^2 \right]\,,
 \end{equation}
where the superscripts $11$ and $22$ in annihilation cross 
sections mean the cross section for $\varphi_i \varphi_i \rightarrow SM$ for $i=1,2$ respectively.
The allowed region of the space of parameters that must be used in solving the Boltzmann equations 
(\ref{2boltz1}) and (\ref{2boltz2}) are $-0.128<\lambda<0$, $\lambda_{H} = \lambda + 0.128$ and 
 $\lambda_i > -1.65 \lambda /\lambda_H$ for $i=1,2$. 

\begin{figure}
\begin{minipage}{0.36\textwidth}
\includegraphics[width=\textwidth,angle =-90]{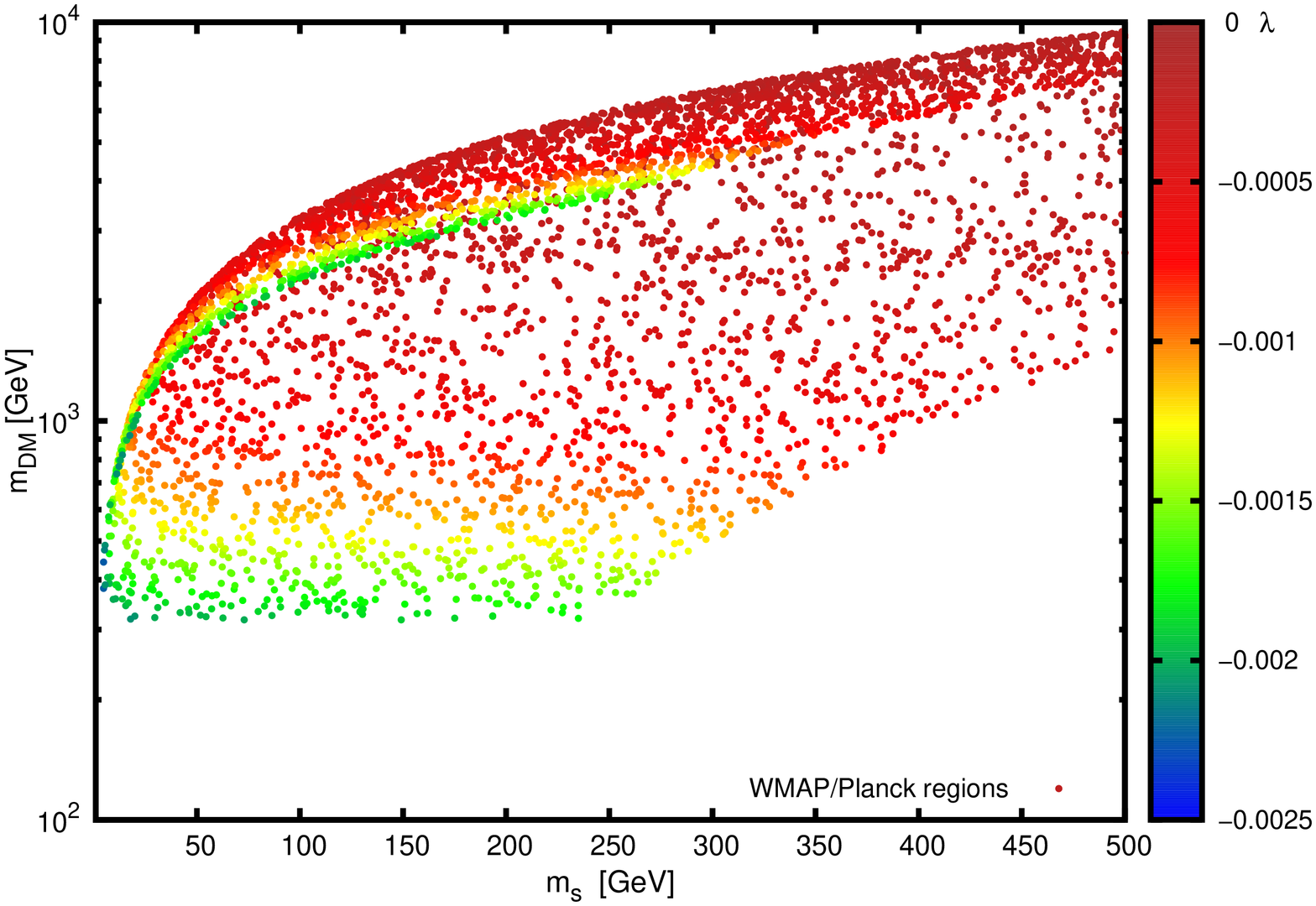}
\end{minipage}
\hspace{2.4cm}
\begin{minipage}{0.36\textwidth}
\includegraphics[width=\textwidth,angle =-90]{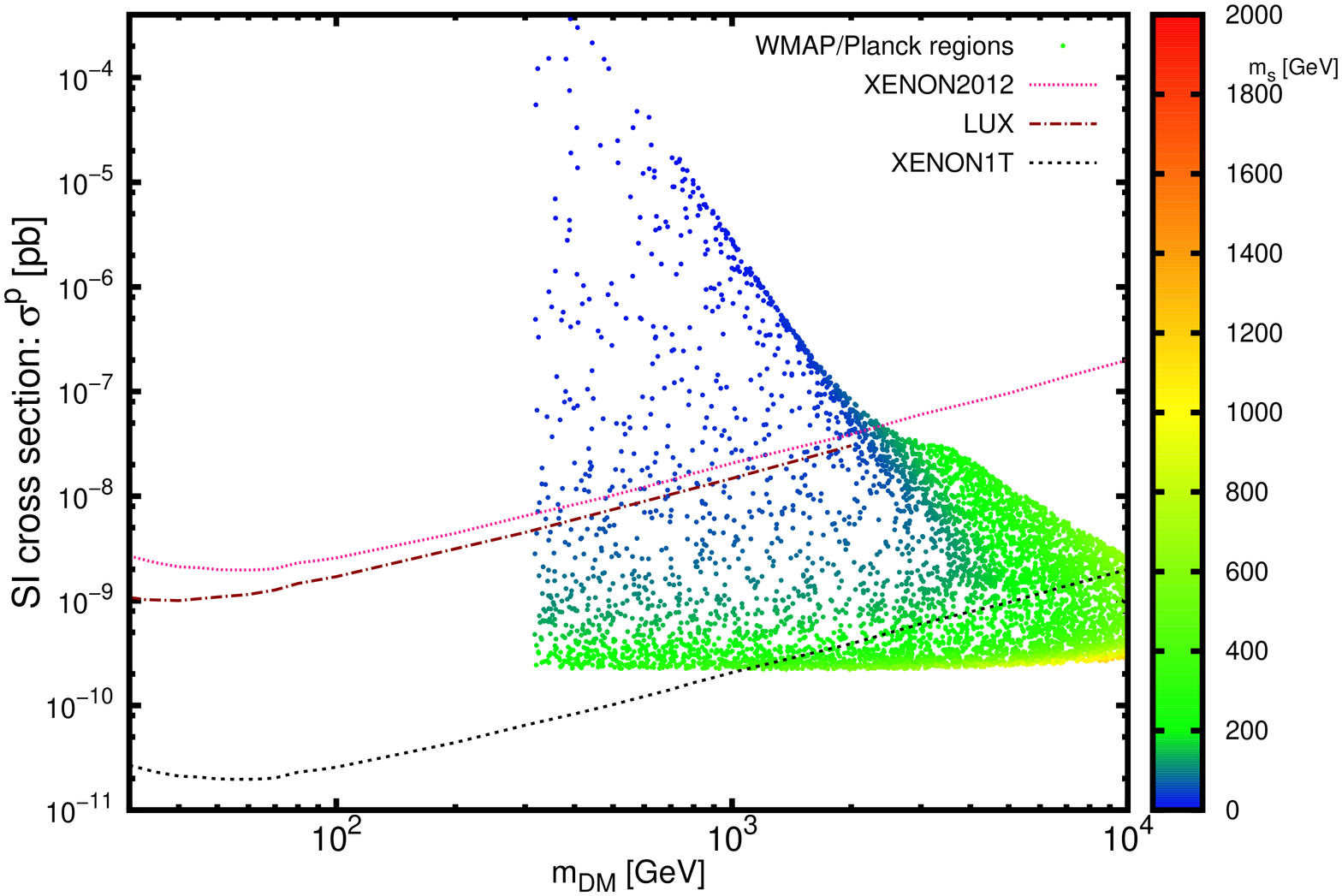}
\end{minipage}
\caption{Direct and Indirect probes for two-component scalar DM in the SISM: {\it Left}) 
The allowed DM mass constrained from DM relic abundance measured by Planck/WMAP against the scalon mass, {\it Right})
The allowed DM mass constrained by Planck/WMAP for relic density and by Xenon100/LUX for the elastic
scattering cross section of DM off the proton.}
\label{2DM}
\end{figure}

\begin{figure}
\begin{minipage}{0.36\textwidth}
\includegraphics[width=\textwidth,angle =-90]{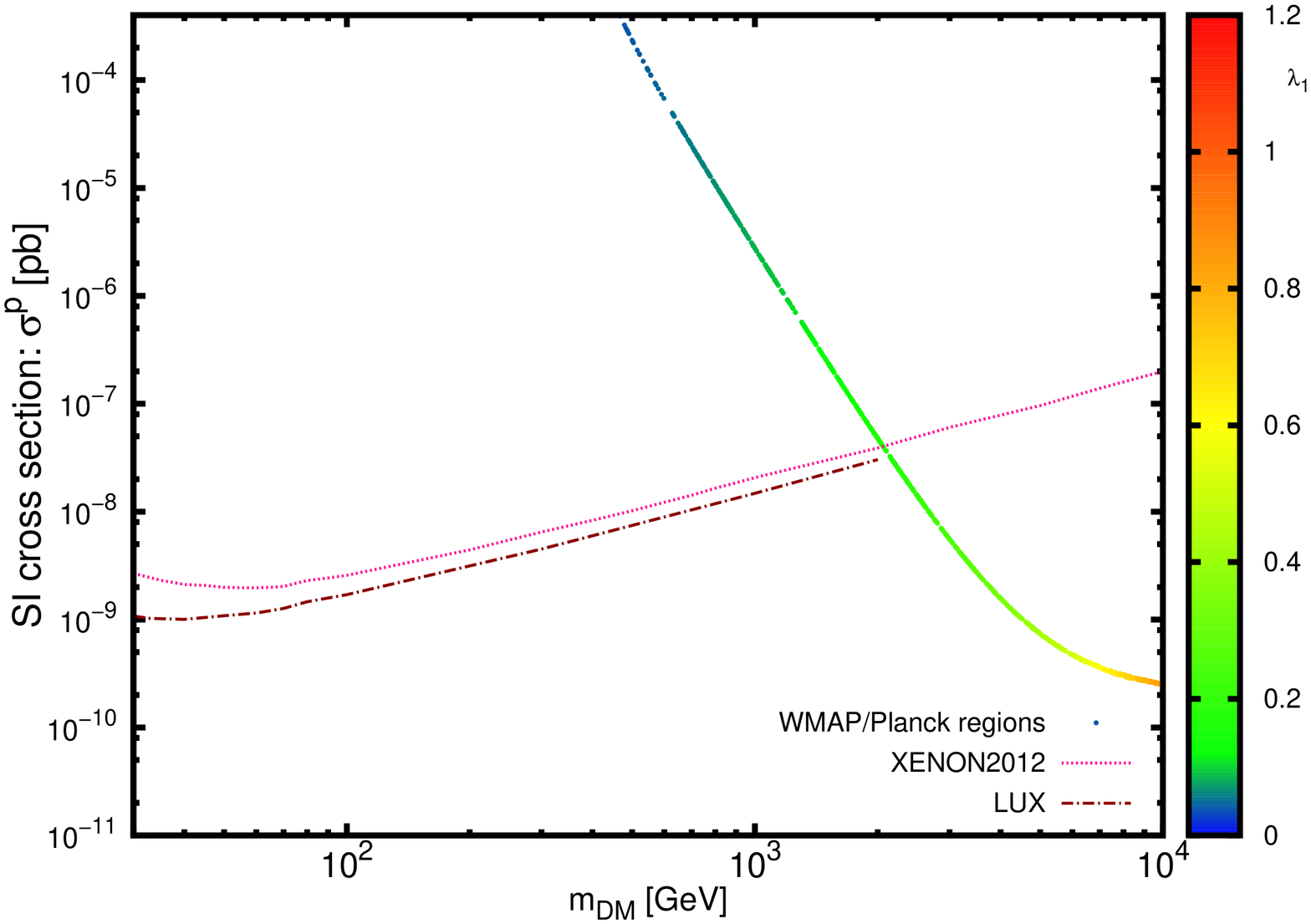}
\end{minipage}
\hspace{2.4cm}
\begin{minipage}{0.36\textwidth}
\includegraphics[width=\textwidth,angle =-90]{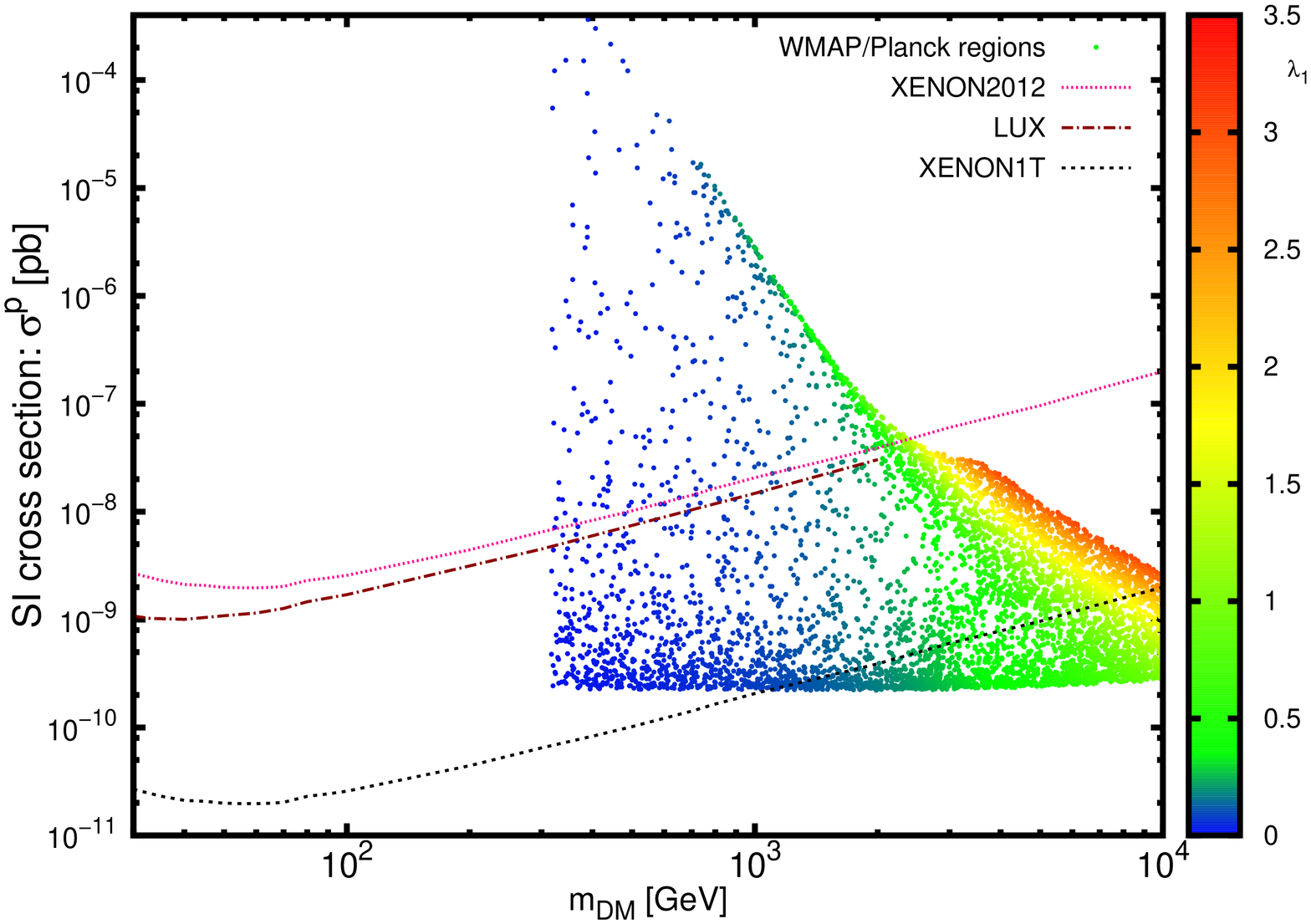}
\end{minipage}
\caption{The $\lambda_1$ dependency of the allowed DM mass constrained by Planck/WMAP for relic density and by Xenon100/LUX for the elastic
scattering cross section of DM off the proton. {\it left}) signle dark matter  
{\it right}) two-component dark matter.}
\label{lambda1}
\end{figure}

The results of the numerical computation for the relic density and the DM elastic scattering cross section 
for two-component scalar
dark matter are shown in Fig. \ref{2DM}. In the right plot of Fig. \ref{2DM} the allowed DM masses
in the viable parameter space of the relic abundance measured by Planck is drawn against the mediator mass. The 
behavior observed in the single scalar DM case is not the same as the two-component DM model, 
i.e. growing the scalon 
mass does not leads necessarily to greater DM mass. In the right plot in Fig. \ref{2DM} the viable parameter
space which evades both relic and direct constraints are shown, We can see a big change 
in the range of the DM mass compared to that of the single DM case analyzed in subsection \ref{SDM}. 
The bound for the mass
 of each scalar DM now is lowered to $m_{\text{DM}} \gtrsim 300$ GeV. Again the non-vanishing scalon mass
 plays an important role in obtaining a viable parameter space. 
 In Fig. \ref{lambda1} we have shown how the DM mass and the associated DM-nuclei cross sections
 constrained by the 
 direct and indirect bounds change with respect to the coupling $\lambda_1$ instead of the coupling 
 $\lambda$ in both single and two-component DM scenarios. In the two-component case, the dependency of the elastic scattering 
 cross section on the coupling $\lambda_2$ is the same as the coupling $\lambda_1$, so we refrain drawing a 
 separate plot for that.

\subsection{Interacting Dark Matter Components}
In the last section the two DM candidates did not have any interaction among themselves. Here we assume an interaction between DM components which respect the scale invariance in the potential (\ref{pot}),
\begin{equation}\label{int}
 V_{\text{int}}=\frac{1}{2}\lambda_{\text{int}}\varphi_1^2 \varphi_2^2\,.
\end{equation}

In our calculations we find out that the presence of the new interacting term (\ref{int}) does not affect significantly
the relic density. More precisely, to see the effect of the coupling $\lambda_{\text{int}}$ we fix all the couplings
in the full lagrangian and change $\lambda_{\text{int}}$ in the interval $-3<\lambda_{\text{int}}<3$. We observe that the changes in
the amount of relic density compared to when we set $\lambda_{\text{int}}=0$ is at most about $\%0.3$ 
in the Planck/WMAP region. 

\section{Discussion}
The minimal supersymmetry standard model is capable of addressing some important drawbacks
in the standard model such as the Higgs hierarchy problem and the issue of dark matter. 
However MSSM due to possessing many free parameters is very difficult to be detected experimentally as
at LHC no evidence for such a theory has been recorded so far. Motivated 
by this fact we are interested in studying the scale invariant standard model (massless-Higgs standard model) 
which is free of hierarchy problem. In SISM the Higgs boson receives mass from the vacuum expectation
value of another scalar called scalon which remains massless classically. However radiative corrections 
gives a small mass to the scalon that is crucial if we want to extend this theory to include DM candidates.
More scalars in the theory then can play the role of DM particle(s). By adding once one scalar and then 
two scalars we have considered the case of single scalar dark matter and two component dark matter in SISM.
In this paper we have examined the SISM whether it can accommodate the problem of 
dark matter as a WIPM in the freeze-out scenario. We observed remarkably that the SISM despite having a narrow 
parameter space already restricted due to the scale invariance is quite successful in overcoming  
the constraints dark matter relic abundance and the direct detection of dark matter elastic scattering 
of nuclei, far enough below the bounds put by Planck and Xenon100 as seen from Figs. \ref{singleDM} and \ref{2DM}.

In the following we discuss briefly the case in which the SISM is extended by a fermionic DM candidate. 
Suppose in addition to Higgs scalar $h$ and the scalon $s$ the theory possesses a Dirac fermion that is
communicating with the SM sector through the Higgs portal by a Yukawa interaction, 
\begin{equation}\label{fer}
  \mathcal{L}_{\text{int}}=  g s \bar{\psi} \psi  
  + g_5 s \bar{\psi} \gamma^5 \psi\,, 
\end{equation}
where $g$ and $g_5$ are the Yukawa couplings. The coefficients in the effective potential (\ref{Veff}) 
in the absence of any DM scalars gets contribution from the fermion DM in the loop, 
\begin{equation}\label{An1}
\begin{split}
 A(\bar{\mathbf{n}})=\frac{1}{64\pi^2 v_{\phi}^2} 
 \bigg[ m_h^4 \left( -\frac{2}{3} 
 + \log{\frac{m_h^2}{v_\phi^2}}\right)
 + 6 m_{W}^4 \left( -\frac{5}{6} 
 + \log{\frac{m_W^2}{v_\phi^2}}\right) 
 + 3 m_Z^4 \left( -\frac{5}{6} 
 + \log{\frac{m_Z^2}{v_\phi^2}}\right)\\
 -12 m_t^4 \left( -1
 + \log{\frac{m_t^2}{v_\phi^2}}\right) 
 - 4 m_\psi^4 \left( -1 
 + \log{\frac{m_\psi^2}{v_\phi^2}}\right) \bigg]\,,
 \end{split}
\end{equation}
and 
\begin{equation}\label{Bn1}
 B(\bar{\mathbf{n}})=\frac{1}{64\pi^2 v_{\phi}^4} 
 \left( m_h^4 + 6m_W^4 + 3m_Z^4 -12 m_t^4 -4m_\psi^4 \right)\,.
\end{equation} 
The mass of the scalon then takes the form 
\begin{equation}\label{m_s1}
 \delta m_s^2 = 2 B v_\phi^2=-\frac{\lambda}{32\pi^2 m_H^2} 
 \left( m_H^4+ 6m_W^4 + 3m_Z^4 -12 m_t^4 -4m_\psi^4\right)\,.
\end{equation}
The coupling $\lambda$ takes only negative values as pointed out in section \ref{prob}. 
It should be noted first that even before adding the DM fermion the scalon mass correction becomes negative 
due to the existence of the heavy top quark  in the loop. So the presence of the DM fermion 
deteriorates the situation. In addition to Higgs scalar and the scalon, it is therefore necessary to
add more bosons (scalars or vector bosons) to the theory. In the presence of further bosons 
the issue of fermionic DM in SISM is at least consistent by construction (see \cite{Benic:2014aga} for an example). 
However 
the positivity of the 
scalon mass restricts strongly the mass of the Dirac fermion $\psi$ and the scalon $s$ remains always very light,
which in turn as discussed in subsection \ref{SDM} the theory might be ruled out by constraints from direct
detection tests.

\section*{Acknowledgment}
We would like to thank A. Pukhov for very useful discussions and many email exchanges on the micrOMEGAs.
K.GH acknowledges Arak University for a grant under the contract 93/4092.
\bibliography{Ref}{}
\bibliographystyle{JHEP}
\end{document}